# Dual-Path Mechanism of Amino Acid Racemization Mediated by Quantum Mechanical Tunneling

Xinrui Yang,[a), b)] Rui Liu,[b)] Ruiqi Xu,[a)] Zhaohua Cui[b)] and Zhigang Wang[a), b)]

[a)] Institute of Atomic and Molecular Physics, Jilin University, Changchun 130012, China.

[b)] Key Laboratory of Material Simulation Methods & Software of Ministry of Education, College of Physics, Jilin University, Changchun 130012, China.

**TOC GRAPHIC:**

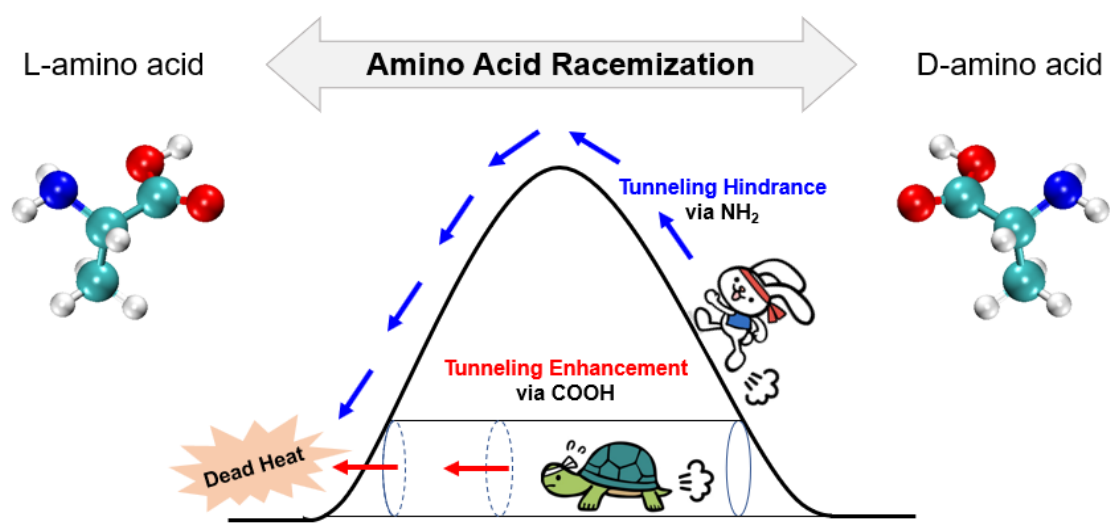

**Abstracts**

    The racemization of amino acids constitutes one of the most elemental and critical reactions, holding primitive significance for understanding the life's origin and maintenance. Nevertheless, its mechanism at the atomic level has been persistently misunderstood for more than a century. In this work, we demonstrate that the racemization of amino acid molecules in aqueous environments can occur simultaneously by two pathways via the carboxyl (COOH) and amino (NH$_2$) groups. Behind this result, the quantum mechanical tunneling (QMT) effect plays a pivotal role, as evidenced by the tunneling hindrance of the NH$_2$ reaction and the tunneling enhancement of the COOH reaction. Notably, the disparity in the QMT effect leads to a crossover between the COOH and NH$_2$ reactions within 200-257 K, such that NH$_2$ reactions dominate at high temperatures and COOH reactions dominate at low temperatures. Our work emphasizes the significance of QMT effect in the racemization of amino acids and therefore introduces a dual-path coexistence mechanism, offering valuable insights into the origin of homochirality in extreme environments of the early Earth.

**Introduction**

    The racemization of amino acids, as a necessary consideration for discussing the origin of homochirality and even life,[1–4] has attracted great interest in multiple fields, such as biochemistry,[5] geological dating,[6–11] drug development,[12] disease research,[13,14] and peptide chemistry.[15,16] An in-depth understanding of its mechanism will be essential for the development of related research. Currently, the generally accepted atomic-level mechanism of amino acid racemization is the ionic model proposed by Neuberger in 1948,[17] i.e., the α-H atom on the chiral C atom is abstracted by free OH$^-$ ions to form a planar carbanion, which in turn adsorbs a H$^+$ from water molecules on the other side. This model can be traced back as far as Dakin's description of the racemization in 1910,[18] and has been widely accepted as the primary mechanism of amino acid

racemization for more than a century. However, this ionic model is only proposed to describe the racemization with "very high concentration of base" as Neuberger claimed,[17] which leads to an inability to explain the experimental results that the rate of racemization at around pH = 7 is essentially unchanged with pH.[19–21] Therefore, this ionic model does not seem to be suitable to describe the amino acid racemization process that occurs commonly in nature.

What exactly is the mechanism of amino acid racemization? Regarding this issue, theoretical simulations of racemization processes have always been of concern due to the serious challenges faced by direct experimental observation for a long time. As early as 2003, the amino acid racemization pathway via amino group ($NH_2$) was first predicted, i.e., the α-H atom is transferred to the amino N atom, which makes the amino acid molecule planar and then undergoes racemize.[22] Around 2010, the α-H atom in amino acid residues was proved to be also transferred to the carbonyl O atom in carboxyl group (COOH) to induce racemization, and further evidence was provided to support the facilitating role of water molecules as a bridge in this process.[23–25] In 2019, it was found that compared to the ionic model, the water-bridge reaction via $NH_2$ is more suitable for describing the racemization process of serine (Ser).[26] As can be seen, the issue of the moment seems to be determining which one of these two models via COOH and via $NH_2$ is more consistent with the experimental results. However, as early as 1980, experiments on dipeptides had already revealed that the racemization rate of COOH-terminal amino acid residues is similar to that of the $NH_2$-terminal amino acid residues, contradicting either of above two models.[27] Considering these results, despite the development of a modeled comprehension regarding the involvement of water molecules, it is imperative to assess whether specific factors may have been inadvertently overlooked in the fundamental physical effects. This oversight potentially underlies the inability of atomic-level models to align with experimental data across various disciplines.

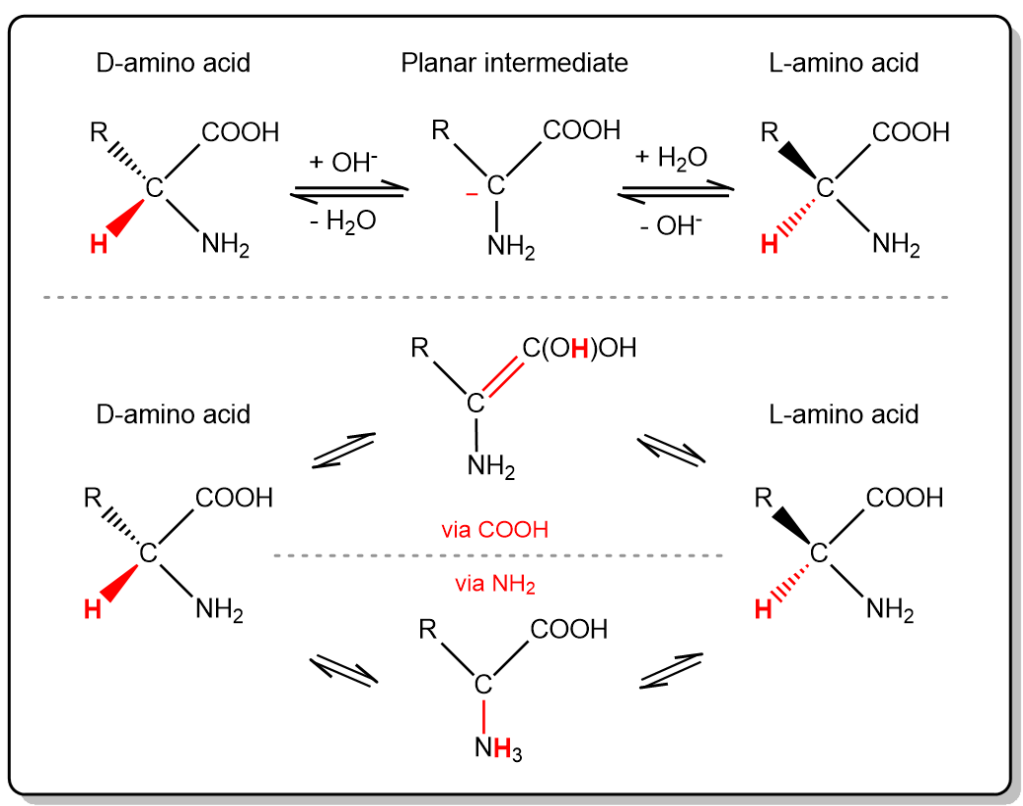

**Scheme 1**. Three models of the amino acid racemization.[24,26,28]

The opportunity to reconsider this mechanism arises from the groundbreaking understanding of the quantum mechanical tunneling (QMT) effect in the chiral inversion process. As early as 1927, the first proposal of quantum tunneling was used to understand the chiral inversion process in chiral molecules, but the importance of QMT in chiral inversion processes has never been demonstrated.[29–31] Until 2023, research has revealed that the QMT effect constitutes the pivotal factor in the racemization of thalidomide,[32] a key molecule that enables people to realize the importance of chirality, thereby revealing a deep physical mechanism behind the chiral inversion. As the significance of QMT in the racemization process has been demonstrated and recognized, the QMT in amino acid racemization needs to be deeply investigated, which may be the ultimate secret code to reveal its atomic-level mechanism.

Against this background, we investigate the racemization processes of typical amino acid molecules, Alanine (Ala), Isoleucine (Ile) and Valine (Val), in aqueous environments. Based on the high-precision potential energy surface (PES), the calculations of the small curvature tunneling (SCT) corrected reaction rate constants clearly show that the amino acid racemization in nature occurs simultaneously by two water-bridge reaction pathways (via COOH and via $NH_2$), in which QMT effect plays an essential role. This leads to the fact that although the barrier height of the COOH reaction (~1.5 eV) is greater than that of the $NH_2$ reaction (~1.2 eV), these reaction rate constants are comparable. Further, the difference in tunneling ability between the two reactions is profoundly shown to originate from the asymmetry of the barriers, rather than from just the barrier height or width, as is briefly assumed in the past. These findings clarify long-standing doubts about the mechanism of amino acid racemization and prove the fundamental significance of QMT in chiral inversion.

## Results and discussion

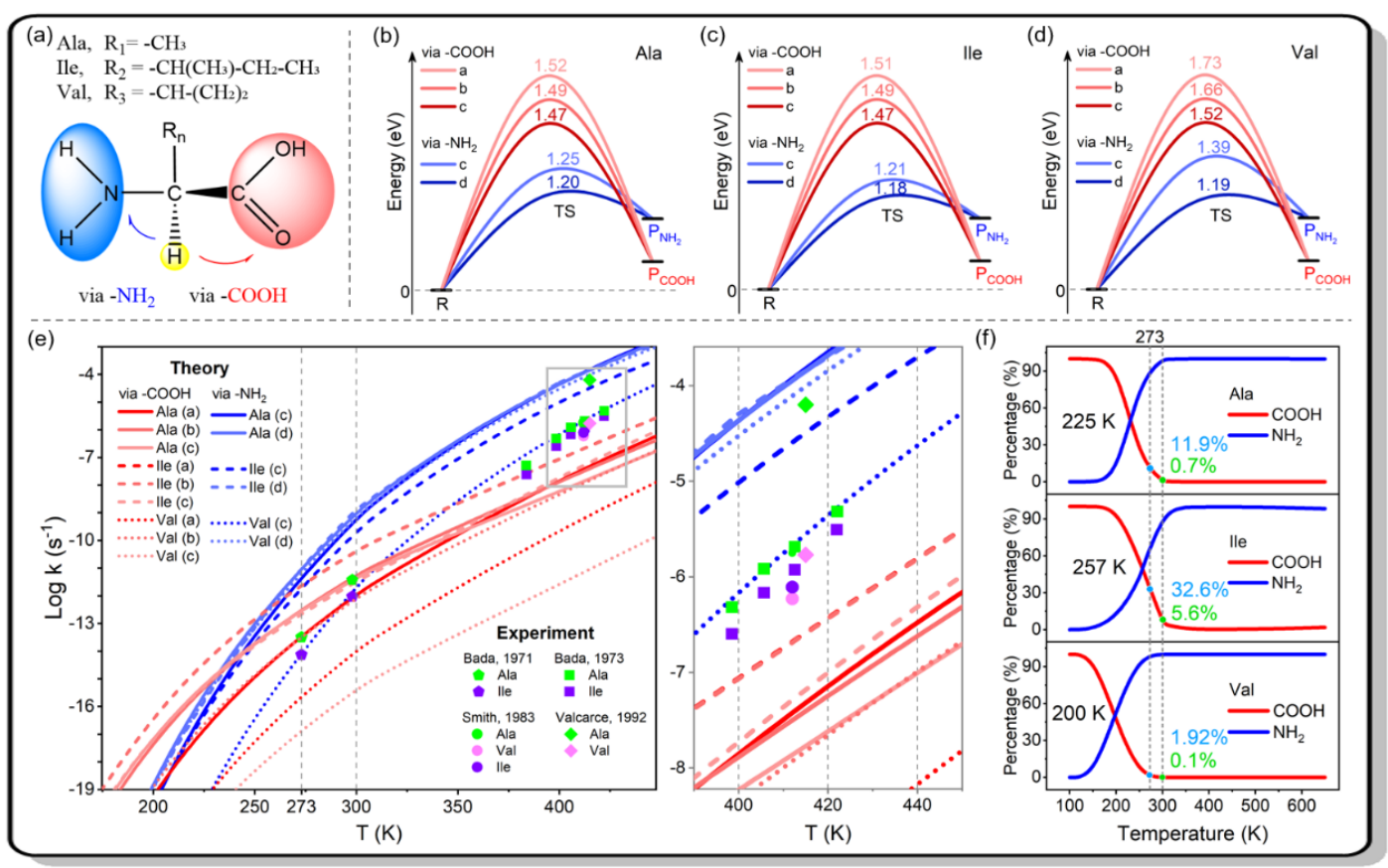

**Figure 1.** (a) Schematic diagram of the two pathways of amino acid racemization (via COOH and via $NH_2$). (b-d) Potential energy surfaces (PESs) of the rate-determining steps for three typical amino acid molecules, Alanine (Ala), Isoleucine (Ile) and Valine (Val). For all reactions, the participation of two water molecules is considered. TS denotes the transition state and R denotes the reactants. $P_{NH2}$ and $P_{COOH}$ represent the products of two $NH_2$ reactions and COOH reactions, respectively. The energy values were all corrected by the zero-point energy. (e) Theoretical predictions in our work and previous experimental observations of the reaction rate constant for the amino acid racemization.[8,20,33,34] The gray square indicates a region that has been enlarged, and the enlarged view of this region is displayed to the right. (f) The ratio of reaction rate constants of two reactions via COOH and via $NH_2$.

To unravel the deep mechanism of amino acid racemization, three typical nonpolar amino acid molecules, Ala, Ile and Val, were considered because of their widely uses in amino acid racemization dating in archaeology.[8,10] For all these amino acid molecules, we have tried to calculate different possible reaction channels and confirm that the rate-determining steps for the reactions are the transfer processes of the α-H atom on the chiral C atom to COOH and $NH_2$.[22,24] A schematic illustrating these rate-determining steps is presented in Figure 1(a). Further, the molecular conformation of each amino acid molecule was carefully considered. Using the dihedral angle of COOH as the coordinate φ$_{COOH}$ and the dihedral angle of $NH_2$ as the coordinate φ$_{NH2}$, 6 stable structures (φ$_{COOH}$, φ$_{NH2}$) were found for each amino acid molecule, (1,1), (1,2), (2,1), (2,2), (3,1),

(3,2) (see Figures S1-3 in the Supporting Information, SI). For each conformation, two water molecules are considered to facilitate the α-H atom transfer process, as previously reported.[24–26] After a refined reaction pathway search on PESs, it is found that for the three amino acid molecules, the (1,1), (2,1), (3,1) conformations can undergo a reaction via COOH, while (3,1), (3,2) can undergo a racemic path via $NH_2$ (see Figures S4-9 in the SI). For convenience, the four configurations (1,1), (2,1), (3,1), (3,2) are referred as a, b, c, d and the 5 reaction processes are labeled as COOH(a), COOH(b), COOH(c), $NH_2$(c), $NH_2$(d), respectively. The aforementioned results concerning structures and reaction pathways lay the groundwork for the subsequent exploration of QMT effect, which will be discussed in detail next.

To begin with, the potential barriers of two water molecules on the 5 reactions of each amino acid molecule were calculated and shown in Figures 1(b-d). The results indicate that the energy barriers associated with the COOH reaction are obviously greater than these in the $NH_2$ reaction. Specifically, for Ala, the energy barriers corresponding to the COOH(a), COOH(b), and COOH(c) reaction channels are 1.52, 1.49, and 1.47 eV, respectively, while the barriers for the $NH_2$(c) and $NH_2$(d) channels are 1.25 and 1.20 eV, respectively. For Ile, the energy barriers of the three reaction channels COOH(a), COOH(b), and COOH(c) are 1.51, 1.49, and 1.47 eV, respectively. The barriers for the $NH_2$(c) and $NH_2$(d) channels are 1.21 and 1.18 eV, respectively. For Val, the energy barriers of the three reaction channels via COOH, namely COOH(a), COOH(b), and COOH(c), are 1.73, 1.66, and 1.52 eV, respectively, and the barriers for the two reaction channels via $NH_2$, namely $NH_2$(c) and $NH_2$(d), are 1.39 and 1.19 eV, respectively. It can be seen that for all three amino acids, the barriers for the reactions via $NH_2$ are smaller than these via COOH, as if to predict that $NH_2$ reaction is more dominant in the two amino acid racemic pathways.

Considering the reaction rate is the most important and identifiable physical quantity in the researches of amino acid racemization, the quantum-corrected reaction rate constants were computed. As shown in Figure 1(e), for each amino acid molecule, the reaction rate constants via COOH and via $NH_2$ exhibit very similar values. For clear presentation, the fastest rate constant for each racemization pathway for each amino acid is taken as the representative rate constant. Specifically, the results show that at 300 K, the racemization rate constants of Ala are $5.66\times10^{-12}$ and $1.03\times10^{-9}$ $s^{-1}$ for the reactions via COOH and via $NH_2$; for Ile, the corresponding racemization rates are $4.28\times10^{-12}$ and $1.28\times10^{-9}$ $s^{-1}$; and for Val, the racemization rates are $8.45\times10^{-13}$ and $7.18\times10^{-10}$ $s^{-1}$ for reactions via COOH and $NH_2$, respectively. Besides, from Figure 1(e), it can be obviously seen that there is an intersection of the reaction pathways via COOH and via $NH_2$ in different amino acids in the 200-300 K region, which reveals a mechanistic change in the amino acid racemization with temperature, i.e., at lower temperatures, the COOH reaction dominates, and at higher temperatures, the $NH_2$ reaction dominates. In other words, it implies the unusual phenomenon that the two reaction pathways coexist in the amino acid racemization process.

More importantly, combing of previous experimental data, it can be found that the data fall right in the middle of the reaction rate constants for the two types of reactions via COOH and via $NH_2$.[8,20,33,34] Considering that rate-determining steps of 5 reactions lie in the transfer of the only α-H atom and differ only in conformation, it suggests that there may be a complex competition between the 5 reactions for each amino acid molecule. In order to describe this complex mechanism process, we calculated the ratio of each reaction in Figure 1(f) based on the following equations in chemical reaction kinetics:

$$\Phi_{COOH} = \frac{\Sigma k_{COOH(i)}}{\Sigma k_{COOH(i)} + \Sigma k_{NH_2(j)}} (i = a, b, c; j = c, d)$$

$$\Phi_{NH_2} = \frac{\Sigma k_{NH_2(j)}}{\Sigma k_{COOH(i)} + \Sigma k_{NH_2(j)}} (i = a, b, c; j = c, d)$$

where $k_{COOH(i)}$ and $k_{NH_2(j)}$ represent the reaction rate constants through the reactions via COOH as well as $NH_2$, respectively. In Figure 1(f), it can be observed that at lower temperatures, the COOH reaction was absolutely dominant. At 100 K, the percentages of COOH reaction in the three amino acids are about 100%. The percentages of COOH reactions gradually decrease with increasing temperature, while the percentages of $NH_2$ reactions gradually increase, with a cross point at 225 K, 257 K, and 200 K for three amino acid molecules Ala, Ile and Val, respectively. The percentages of $NH_2$ reactions increase with temperature. There is a clear coexistence and competition between the two reactions at 273 K, at which time the proportions of COOH are 11.9%, 32.6%, and 1.92% for Ala, Ile and Val. When the temperature is 300 K, the proportions of COOH reactions are reduced but still present, which are at 0.7%, 5.6%, and 0.1%, respectively. This reveals the competing mechanisms of the two amino acid racemic models with temperature region. When the temperature is high (> 300 K), only $NH_2$ reaction exists and when the temperature is low (< 200 K), only COOH reaction exists. When the temperature is between 200K and 300K, both reactions coexist and compete with each other. Even at 300 K, it is also confirmed that the two mechanisms coexist.

Interestingly, at higher temperatures (> 300 K), thermodynamic effect plays a major role and the reaction path via $NH_2$ dominates, which is consistent with the results of the PESs. At lower temperatures (< 200 K), via COOH dominates instead. This result, which is counterintuitive to the potential energy surface results, seems to predict a stronger QMT effect for the COOH path. To visualize the differences in QMT effect between the different reaction paths, the classical reaction rates (black lines) without QMT effect are shown in Figure 2, and the tunneling correction factors $\kappa(T)$ on the different reaction paths are counted. At 300 K, the $\kappa(T)$ for the reaction path via COOH is larger, with about $10^4$ times, which indicates a stronger tunneling ability. The tunneling correction factor for the reaction path via $NH_2$ is smaller, only a range of 2-3, which indicates its weaker tunneling ability. The difference in tunneling ability explains the paradoxical phenomenon that results from a large difference in the potential barriers of the two reactions but similar reaction rates. In other words, the relationship and nature of the two amino acid racemization mechanisms can only be understood if the QMT effect is clearly considered.

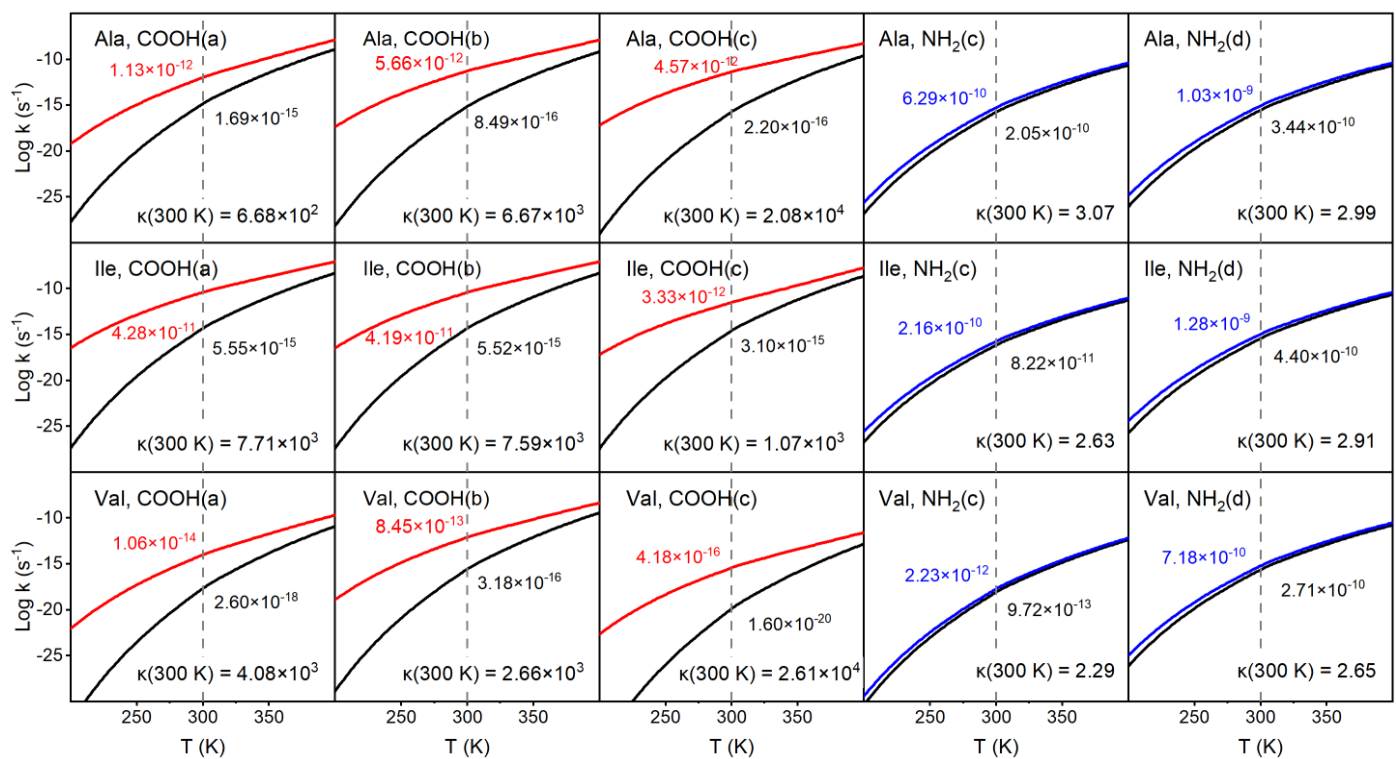

**Figure 2** Theoretically calculated reaction rate constants for COOH(a), COOH(b), COOH(c), $NH_2$(c), $NH_2$(d). The red and black lines represent the quantum-corrected and classical reaction rate constants for the paths via COOH and via $NH_2$, respectively. The black lines are classical reaction rates.

In order to reveal the origin of the difference in quantum tunneling ability between the two reaction paths, we have calculated the tunneling probability ($P_{tunneling}$) by means of the Wenzel-Kramers-Brillouin (WKB) approximation.[35,36]

$$P_{tunneling} = Exp\left[-\frac{2}{\hbar}\int_{x_1}^{x_2}\sqrt{2m(V(x)-E)}\,dx\right]$$

where V is the PES and $E_{in}$ is the input energy. $P_{tunneling}$ for the 15 reaction pathways of the three amino acids were calculated and displayed in Figures 3(a-c). The results show that with the increase of $E_{in}$, the $P_{tunneling}$ increases exponentially. At the same $E_{in}$, the $P_{tunneling}$ of $NH_2$ reaction is generally greater compared to the COOH reaction. For example, when the $E_{in}$ is 1.2 eV, the $P_{tunneling}$ of COOH(a), COOH(b), COOH(c), $NH_2$(c), and $NH_2$(d) are $1.89\times10^{-13}$, $1.65\times10^{-14}$, $2.16\times10^{-14}$,

$5.61×10^{-10}$, $3.41×10^{-3}$. The $P_{tunneling}$ of $NH_2(d)$ is the largest, which is $1.80×10^{10}$, $2.12×10^{11}$, $1.58×10^{11}$, and $6.08×10^{6}$ times higher than the other four reactions COOH(a), COOH(b), COOH(c), and $NH_2(c)$, respectively. This result demonstrates stronger tunneling for the $NH_2$ reaction at the same input energy, which is seemingly inconsistent with the results speculated in the former section. This is due to the fact that based on the definitions, $P_{tunneling}$ can only describe the tunneling capacity at a particular $E_{in}$, whereas the $\kappa(T)$ seems to describe the overall tunneling capacity. Thus, a new question needs to be considered: how should the results of the $\kappa(T)$ be interpreted based on the results of the $P_{tunneling}$?

Based on the definition of $\kappa(T)$, the numerical relationship between $\kappa(T)$ and $P_{tunneling}$ in variational transition state theory (VTST) can be known as follows:[28]

$$\kappa(T) = \frac{1}{k_B T} \int_{E_0}^{+\infty} P_{tunneling}(E) \cdot e^{-(E-V^{AG})/k_B T} dE$$

where $k_B$ is Boltzmann's constant and $V^{AG}$ stands for the maximum of the vibrationally adiabatic ground-state potential energy. From this formula, it suggests that the $\kappa(T)$ are complex integral functions whose integral term is the product of the $P_{tunneling}$ and the energy probability based on the Boltzmann distribution, in which higher energies correspond to lower probabilities of molecular conformations. It means that when the input energy is lower, the weighting of its $P_{tunneling}$ is more important, which is determined using the color shades in Figures 3(a-c). Although the $P_{tunneling}$ of $NH_2$ reaction is even larger than that of COOH reaction, it is clear that the overall integral area of COOH reaction is larger than that of $NH_2$ reaction, which makes the COOH reaction a stronger tunneling ability. The reaction paths obtained by the intrinsic reaction coordinate (IRC) in Figures 3(d-f) clearly indicate that the difference in the integral area of the two reactions lies in the asymmetry of the PESs due to the directionality of the tunneling, which results in the tunneling hindrance of the $NH_2$ reaction, in contrast to the tunneling enhancement of the COOH reaction.

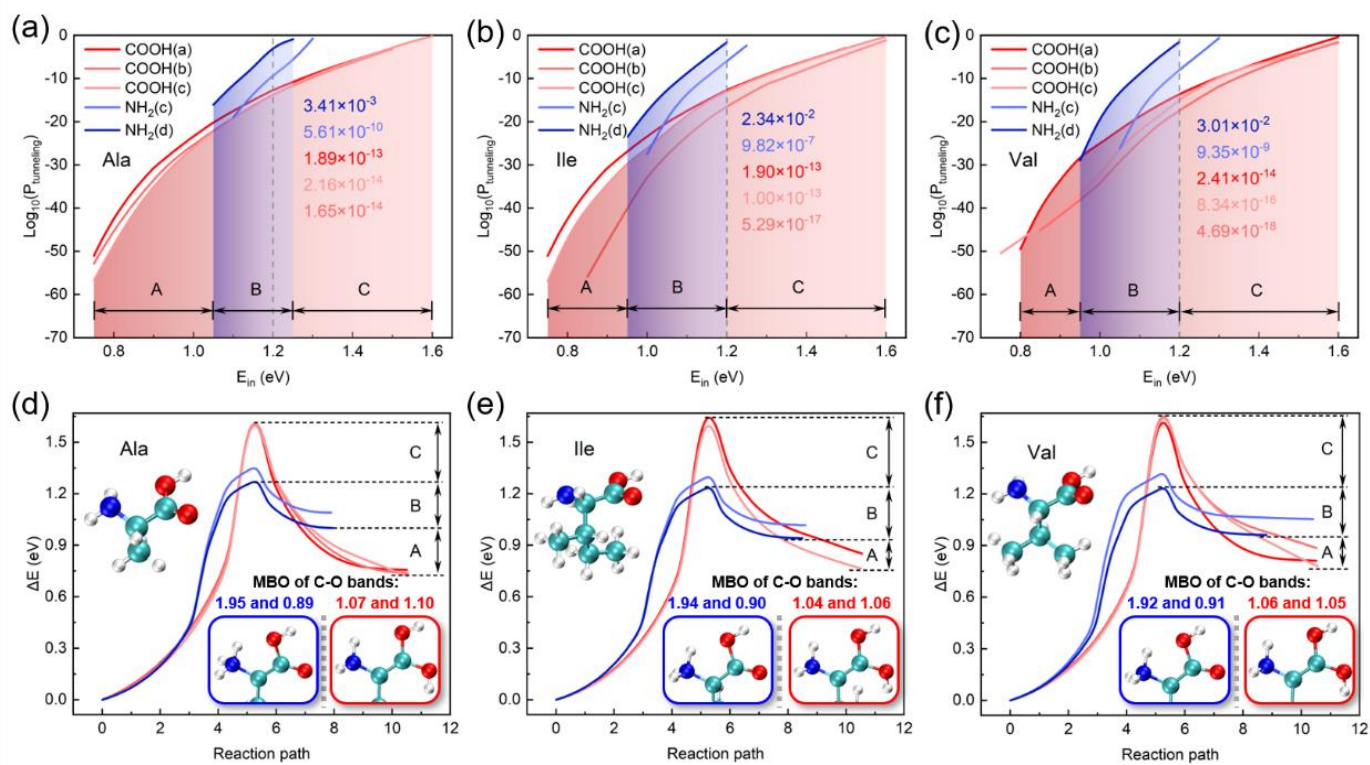

**Figure 3 (a-c)** Tunneling probability ($P_{tunneling}$) of all reactions in this research. The grey dashed line represents that input energy ($E_{in}$) is 1.2 eV. **(d-f)** PESs of all reactions achieved by the intrinsic reaction coordinate (IRC) reaction path. The unit of reaction path is a.u. The locally enlarged molecular structures represent the different products of the two reactions via COOH and via $NH_2$. The blue and red numbers indicate the Mayer bond order (MBO) of the C-O bonds in products.

Based on the above discussion of tunneling probabilities, it can be confirmed that the main factor contributing to the difference in tunneling ability between the two racemic mechanisms for amino acids is the asymmetry of the potential barriers, rather than the barrier width or the barrier height, as is commonly assumed.[37] In order to quantitatively characterize the asymmetry of the reaction process, we first proposed the concept of the asymmetric degree (η), which is defined as

shown below:

$$\eta = \frac{E_P - E_R}{E_{TS} - E_R}$$

where $E_R$ and $E_P$ represent the energies of the reactant and product, respectively, and $E_{TS}$ represents the energy of the transition state. As η tends to 0, it represents a more symmetric PES; while η tends to 1, it represents a more asymmetric PES. Based on this definition, the corresponding η values for the three amino acids were calculated and displayed in Table 1. It can be clearly seen that the η for the COOH reaction is about 0.5, while the η for the $NH_2$ reaction is about 0.8. This indicates that the $NH_2$ reaction potential is more asymmetric, whereas the COOH reaction has a more symmetric potential, which explains the difference in tunneling ability discussed above.

**Table 1** Relative energies of the product and transition state structures relative to the reactants (eV) and the potential asymmetry η corresponding to each path.

|     | Path | $E_P$-$E_R$ | $E_{TS}$-$E_R$ | η | Path | $E_P$-$E_R$ | $E_{TS}$-$E_R$ | η |
| --- | --- | --- | --- | --- | --- | --- | --- | --- |
|     | COOH(a) | 0.76 | 1.61 | 0.472 | $NH_2$(c) | 1.09 | 1.35 | 0.807 |
| Ala | COOH(b) | 0.74 | 1.61 | 0.460 | $NH_2$(d) | 1.00 | 1.27 | 0.787 |
|     | COOH(c) | 0.72 | 1.60 | 0.450 |     |     |     |     |
|     | COOH(a) | 0.85 | 1.65 | 0.515 | $NH_2$(c) | 1.02 | 1.30 | 0.785 |
| Ile | COOH(b) | 0.85 | 1.64 | 0.518 | $NH_2$(d) | 0.94 | 1.23 | 0.764 |
|     | COOH(c) | 0.77 | 1.59 | 0.484 |     |     |     |     |
|     | COOH(a) | 0.81 | 1.61 | 0.503 | $NH_2$(c) | 1.05 | 1.31 | 0.802 |
| Val | COOH(b) | 0.89 | 1.65 | 0.539 | $NH_2$(d) | 0.96 | 1.23 | 0.780 |
|     | COOH(c) | 0.78 | 1.64 | 0.476 |     |     |     |     |

Furthermore, a more in-depth question can be raised: why is the asymmetry of the PESs of the two reaction processes so different? Obviously, this question can be answered by understanding of the behavior of electrons. In the two reactions via COOH and via $NH_2$, the α-H atoms are both transferred and the chiral C atoms undergo transformation from the original $sp^3$ hybridization to a $sp^2$ hybridization, leading to the similar structures of the products. In order to distinguish the minor differences in the electronic structures of the two reaction products quantitatively, Mayer bond order (MBO) analysis of the C-O bonds in products is shown in Figures 3(d-f). For Ala, in the $NH_2$ reaction product, the lower MBO between C atom and hydroxyl O atom (0.89) compared to that between C atom and carbonyl O atom (1.95) suggests that p-π conjugation of the C=O bond hinders p-p conjugation with the hydroxyl group. In contrast, the COOH reaction product exhibits higher and comparable MBOs between C atom and the two hydroxyl O atoms (1.07 and 1.10), indicating enhanced p-p conjugation between the hydroxyl groups. The distinct conjugation behavior emphasizes the increased structural stability of the COOH reaction product, due to unobstructed p-p conjugation between hydroxyl groups, unlike the $NH_2$ reaction product where C=O p-π interaction interferes. Thus, the behavior of the electrons leads to an asymmetry of the PES, which ultimately affects the QMT effect in the reaction process.

## Summary

In summary, our results established the pivotal role of the QMT effect in amino acid racemization, thereby confirming the coexistence of dual-path mechanism within this process. These two pathways exhibit marked disparities in their tunneling capabilities, specifically manifesting as tunneling hindrance for $NH_2$ reaction and tunneling enhancement for COOH reaction. Notably, the discrepancy in QMT behavior is attributed to the asymmetry of the PES induced by the conjugation effect, rather than variations in barrier height or width. Moreover, we demonstrated that these two mechanisms are temperature-dependent: at higher temperatures (> 300 K), the $NH_2$ reaction prevails, whereas at lower temperatures (< 200 K), the COOH reaction dominates. At 273 K and even 300 K, although the COOH reaction is prominent, the $NH_2$ reaction remains significant and non-negligible, indicating that natural amino acid racemization is characterized by the coexistence of both mechanisms, rather than a single pathway. These findings elucidate the temperature-dependent variations in amino acid racemization rates, offering mechanistic insights into the impact of temperature on amino acid racemization dating and potentially contributing to our understanding of the origins of homochirality in the extreme conditions of Earth's early environment.

## ACKNOWLEDGMENT

This work was supported by the National Natural Science Foundation of China (grant numbers 11974136). Z. Wang also acknowledges the assistance of the High-Performance Computing Center of Jilin University and National Supercomputing Center in Shanghai.

# Supporting Information for

# Dual-Path Mechanism of Amino Acid Racemization Mediated by Quantum Mechanical Tunneling

Xinrui Yang,[a), b)] Rui Liu,[b)] Ruiqi Xu,[a)] Zhaohua Cui[b)] and Zhigang Wang[a), b)]

[a)] Institute of Atomic and Molecular Physics, Jilin University, Changchun 130012, China.

[b)] Key Laboratory of Material Simulation Methods & Software of Ministry of Education, College of Physics, Jilin University, Changchun 130012, China.

*Corresponding author. E-mail: wangzg@jlu.edu.cn (Z. W.)

## Computational Method

In this study, to avoid the effect of the polarity of amino acids, three typical nonpolar amino acid molecules, Alanine (Ala), Isoleucine (Ile) and Valine (Val), were considered. The density functional theory (DFT) method combined with reaction rate and tunneling probability calculations was used to investigate the reaction pathways. All relevant structures of thalidomide, encompassing reactants, transition states (TSs), and products, were optimized at the B3LYP-D3 level with the 6-311++g(2df,2pd) basis set using Gaussian 09 software.[1,2] Meanwhile, a third order dispersion correction (D3-BJ) was used to account for the effect of intermolecular hydrogen bonds.[3] During the conformational search of the three amino acids, to reduce variables and ensure a thorough exploration, the R-group configurations of the amino acids were frozen, while the dihedral angles between -COOH, -NH$_2$, and the chiral carbon atom were varied for the search. Based on the optimized conformations of the isolated amino acid molecules, water molecules were added and the system was fully re-optimized. During the search for TSs, potential energy surfaces (PESs) were validated through the intrinsic reaction coordinate (IRC) method.[4,5] In this study, more than 100 points were obtained on each side of the transition state (TS), with calculations performed. Additionally, the calculation and analysis of Mayer bond orders were performed using the Multiwfn software.[6,7]

Reaction rate constants were determined using the POLYRATE 17 package,[8] interfacing with GAUSSRATE 17-B and Gaussian 09.[9] Classical rate constants, excluding tunneling effects, were calculated based on canonical variational transition state theory (CVT).[10] Multidimensional tunneling was assessed using the small curvature tunneling (SCT) approach.[11,12] All rate calculations were performed at the B3LYP-D3(BJ)/6-311++g(2df,2pd) level, with force constant matrices and harmonic vibrational frequencies determined at the same level. The frequency correction factor was also considered.[13] A step size of 0.001 amu^(1/2) bohr was used for computing the minimal energy path (MEP), which was found to be sufficient for detailed characterization of the reaction pathways in this investigation.

The Wentzel-Kramers-Brillouin (WKB) approximation is used for α-H atom transfer processes that are reduced to one-dimensional (1D) coordinates without considering the coherence as the following formula,[14,15]

$$P_{tunneling} = Exp\left[-\frac{2}{\hbar}\int_{x_1}^{x_2}\sqrt{2m(V(x)-E)}\,dx\right]$$

where $\hbar$ is the reduced Planck constant. $V(x)$ represents the PES function with the coordinate $x$ as the variable, $x_1$ and $x_2$ is the two coordinates when $V(x)$ and $E$ are equal. IRC connecting reactants, transition state (TS) structures and products is used to describe the steepest descent path of transition state in mass weight coordinates based on the lowest energy principle. The mass weight coordinates of x axis can be written as $r(i,x) = \sqrt{m(i)} \cdot R(i,x)$ in Cartesian coordinate system. IRC simplifies the complex process with $3N - m$ degrees of freedom into a one-dimensional (1D) reaction process, so WKB approximation can be applied. In order to better describe the PESs along IRC, eighth-order Gaussian Fitting is used to fit the function. The fitting function can be written as:

$$f(x) = \sum_{n=1}^{8}\left(a_n \times e^{-((x-b_n)/c_n)^2}\right)$$

where $a_n, b_n, c_n$ are the Gaussian fitting parameters. It is important to note that the zero-point energy is not included in the calculation of the tunneling probability, but is considered as an additional energy, to provide a clearer understanding of the effect of the zero-point energy on the tunneling probability. It should be also emphasized that the zero-point energy considered at this point may not correspond to all vibrational modes of the system, but rather to the vibrational modes corresponding to the imaginary frequency of the transition state in the proton transfer process, so this may be smaller than the imagined value.

# Figures

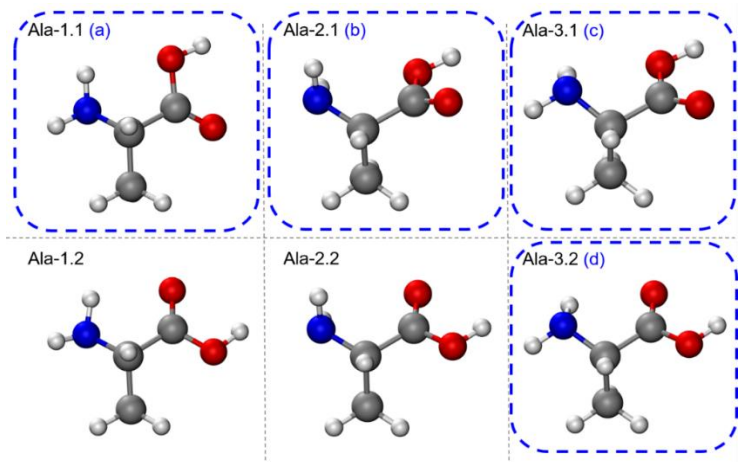

**Figure S1.** The six structures of the Alanine (Ala) molecule.

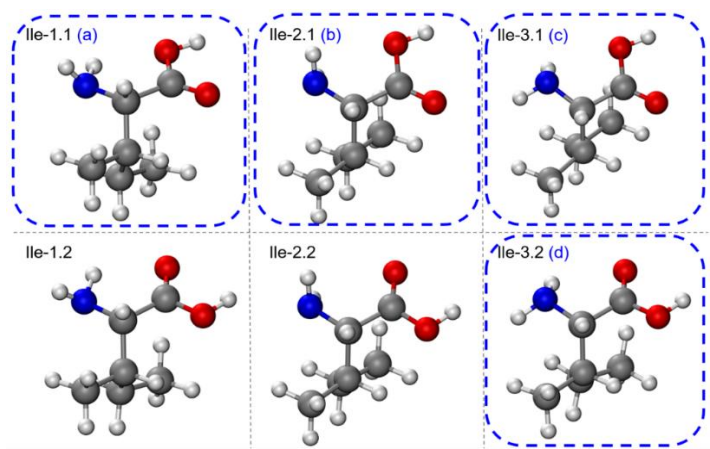

**Figure S2.** The six structures of the Isoleucine (Ile) molecule.

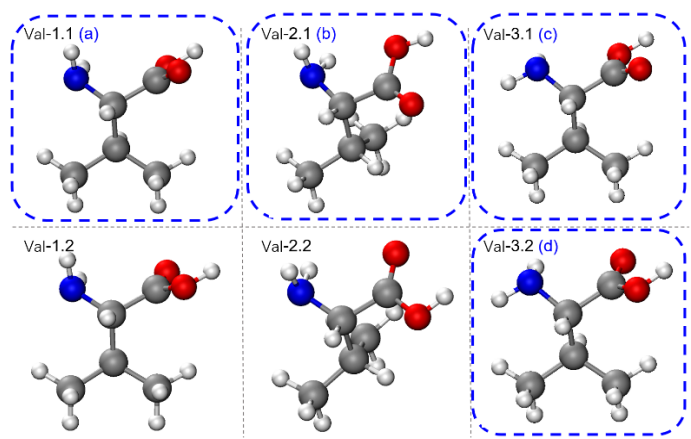

**Figure S3.** The six structures of the Valine (Val) molecule.

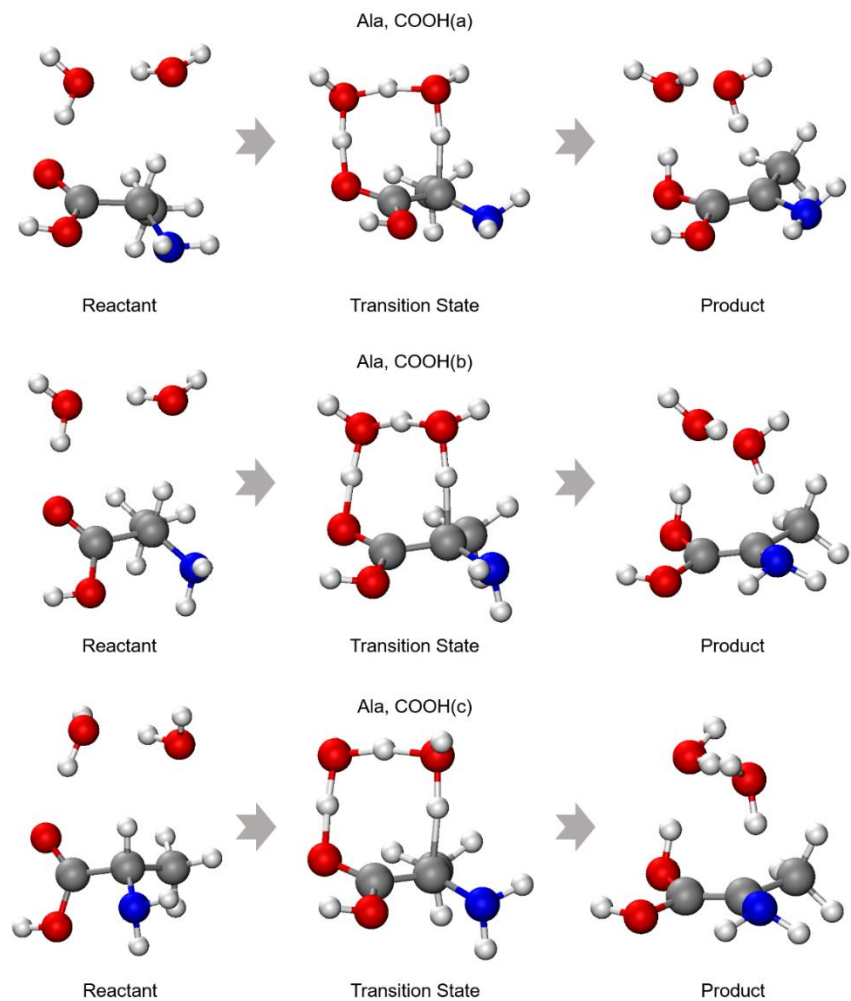

**Figure S4.** The three COOH reactions, COOH(a), COOH(b) and COOH(c), of the Ala molecule with two water molecules.

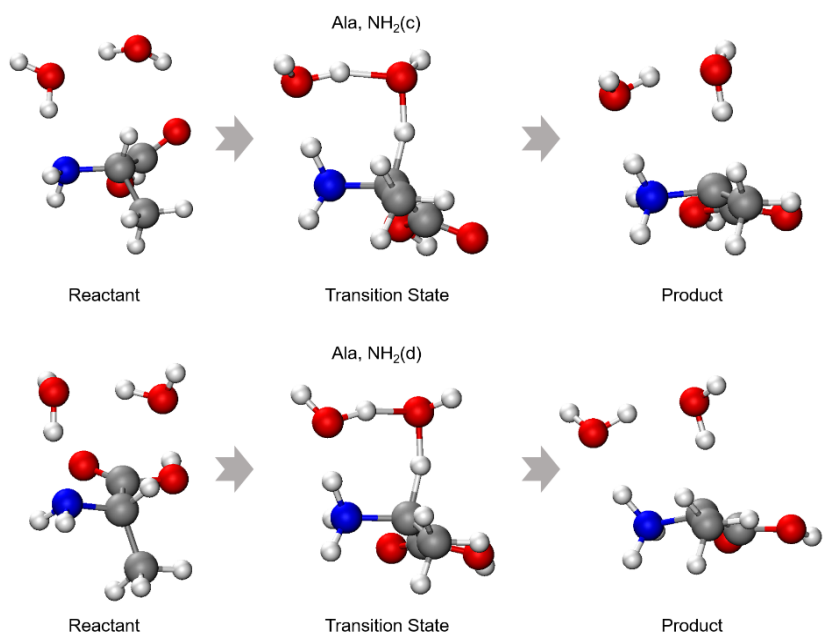

**Figure S5.** The two $NH_2$ reactions, $NH_2$(c) and $NH_2$(d), of the Ala molecule with two water molecules.

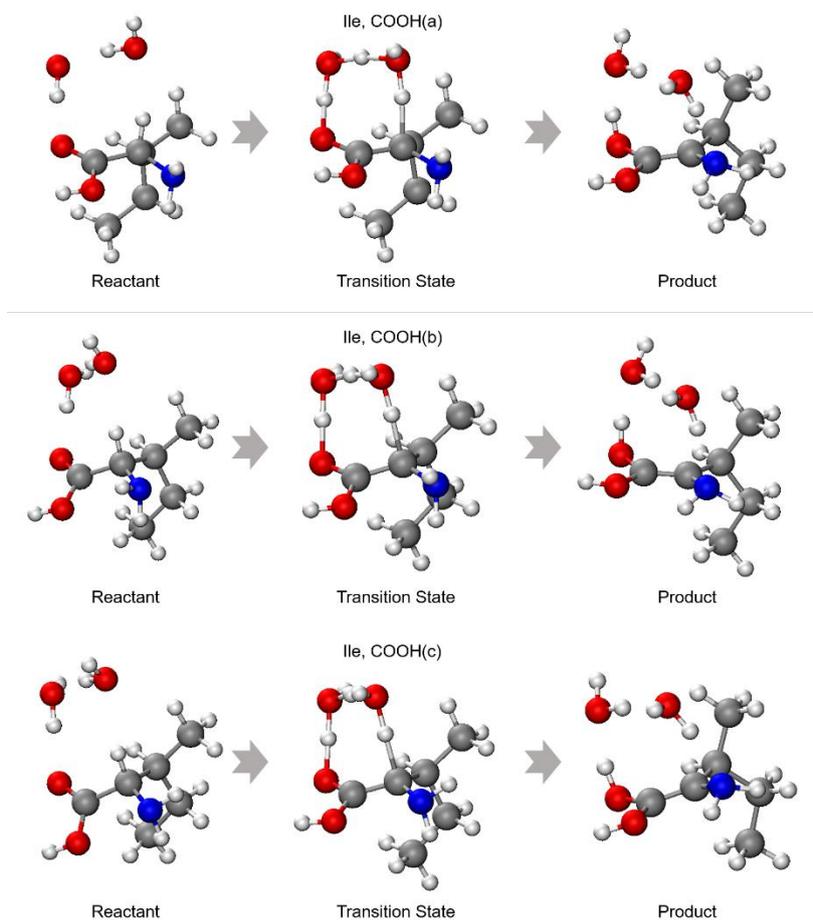

**Figure S6.** The three COOH reactions, COOH(a), COOH(b) and COOH(c), of the Ile molecule with two water molecules.

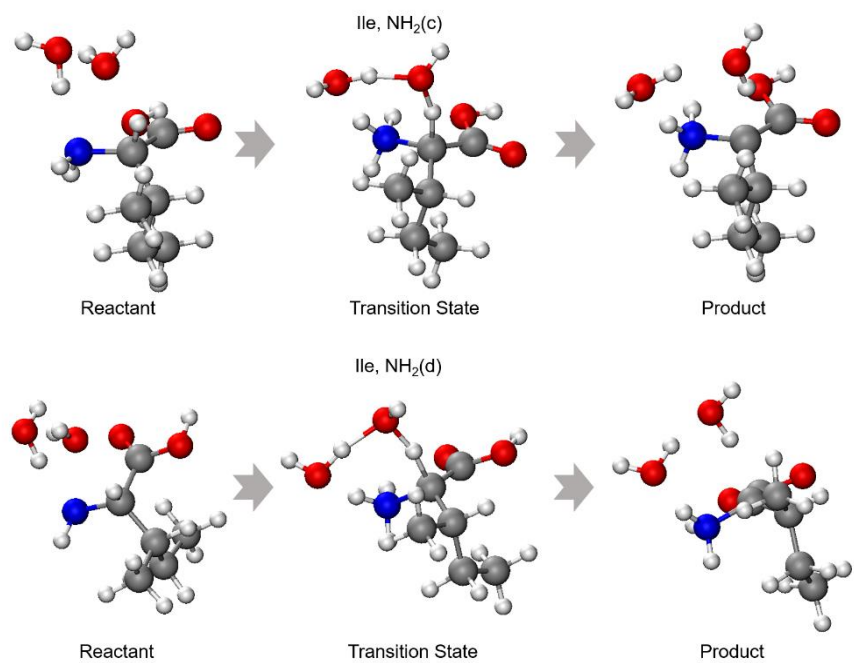

**Figure S7.** The two $NH_2$ reactions, $NH_2$(c) and $NH_2$(d), of the Ile molecule with two water molecules.

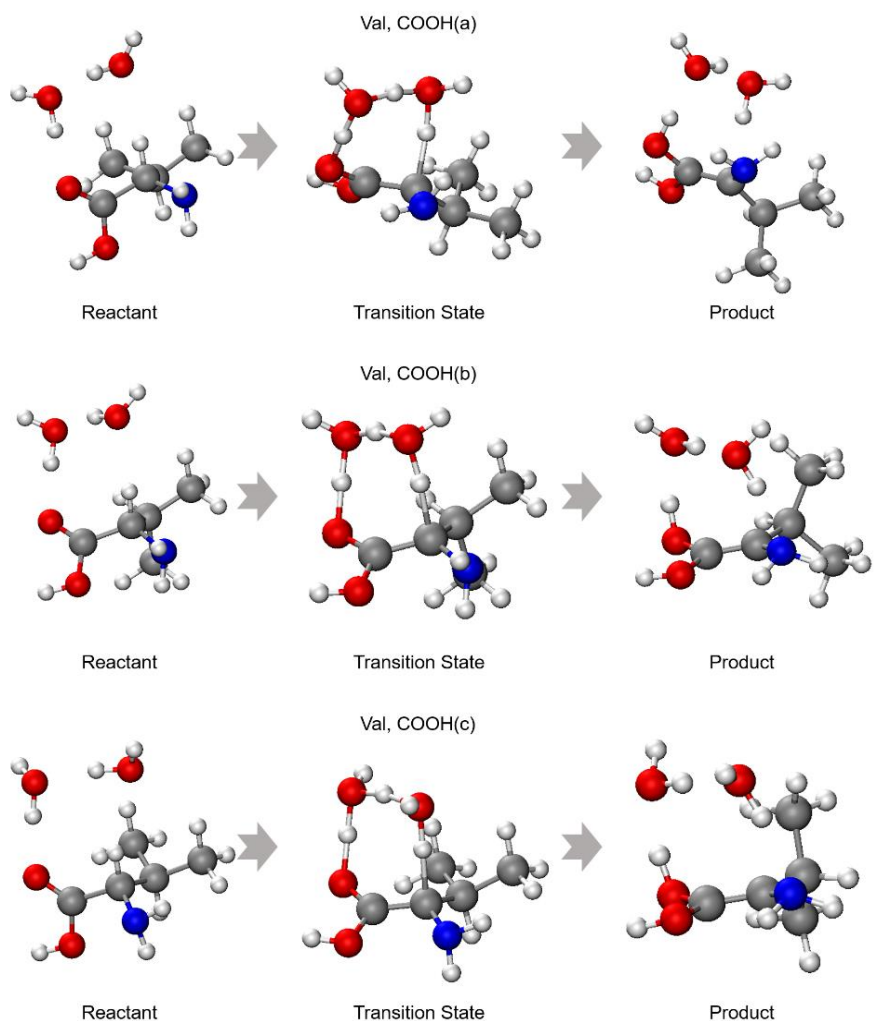

**Figure S8.** The three COOH reactions, COOH(a), COOH(b) and COOH(c), of the Val molecule with two water molecules.

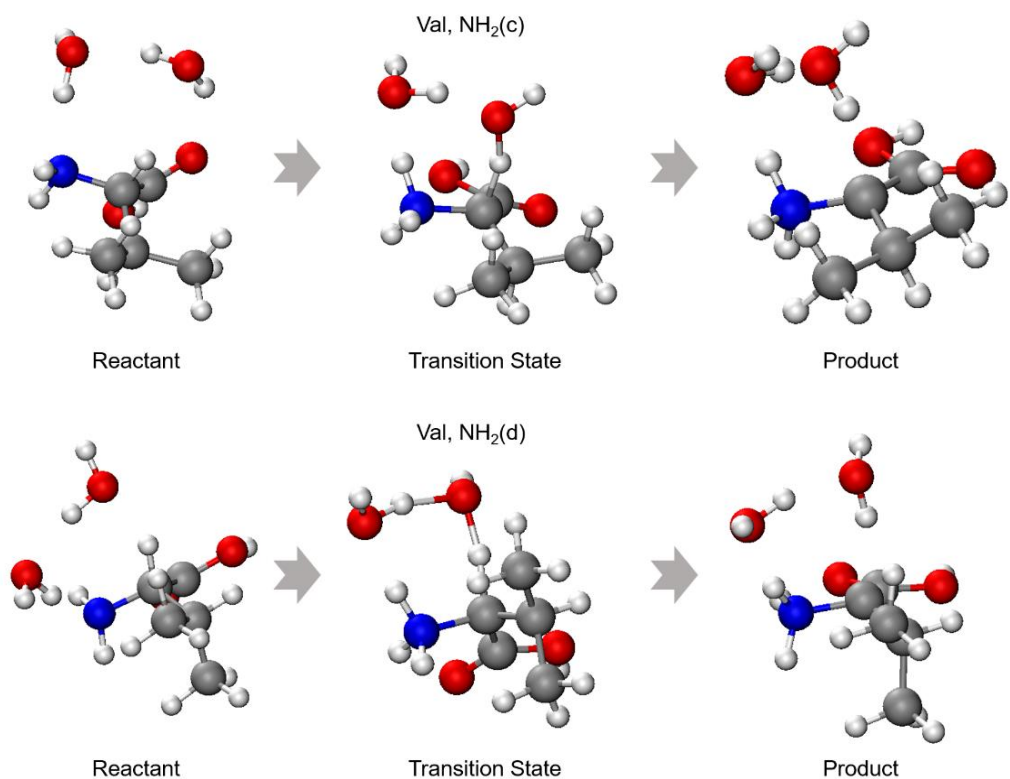

**Figure S9.** The two $NH_2$ reactions, $NH_2$(c) and $NH_2$(d), of the Val molecule with two water molecules.